\documentclass[12pt]{article}
\usepackage{amsfonts}
\usepackage{amsmath}
\usepackage{cite}

\csname @addtoreset\endcsname{equation}{section}
\newcommand{\eq}{\begin{equation}}
\newcommand{\feq}{\end{equation}}
\newcommand{\eqn}{\begin{eqnarray}}
\newcommand{\feqn}{\end{eqnarray}}
\newcommand{\arr}{\begin{eqnarray*}}
\newcommand{\farr}{\end{eqnarray*}}

\newcommand{\sw}{\stackrel{\star}{\wedge}}

\newcommand{\At}{{\tilde A}}
\newcommand{\bt}{{\tilde b}}
\newcommand{\gt}{{\tilde g}}
\newcommand{\tl}{{\tilde{\lambda}}}
\newcommand{\tc}{{\tilde{\chi}}}

\newcommand{\td}{{\tilde{\delta}}}
\newcommand{\hd}{{\hat{\delta}}}
\newcommand{\lp}{\left(}
\newcommand{\rp}{\right)}

\font\mybb=msbm10 at 12pt
\def\bb#1{\hbox{\mybb#1}}

\def\bC {\bb{C}}

\begin{document}

\begin{titlepage}
\begin{flushright}
IFUM-704-FT\\
hep-th/0201103
\end{flushright}
\vspace{.3cm}
\begin{center}
\renewcommand{\thefootnote}{\fnsymbol{footnote}}
{\Large \bf Noncommutative Einstein-AdS Gravity in three Dimensions}
\vskip 15mm
{\large \bf {S.~Cacciatori$^{1,3}$\footnote{cacciatori@mi.infn.it},
D.~Klemm$^{2,3}$\footnote{dietmar.klemm@mi.infn.it},
L.~Martucci$^{2,3}$\footnote{luca.martucci@mi.infn.it}
and D.~Zanon$^{2,3}$\footnote{daniela.zanon@mi.infn.it}}}\\
\renewcommand{\thefootnote}{\arabic{footnote}}
\setcounter{footnote}{0}
\vskip 10mm
{\small
$^1$ Dipartimento di Matematica dell'Universit\`a di Milano,\\
Via Saldini 50, I-20133 Milano.\\

\vspace*{0.5cm}

$^2$ Dipartimento di Fisica dell'Universit\`a di Milano,\\
Via Celoria 16, I-20133 Milano.\\

\vspace*{0.5cm}

$^3$ INFN, Sezione di Milano,\\
Via Celoria 16,
I-20133 Milano.\\
}
\end{center}
\vspace{2cm}
\begin{center}
{\bf Abstract}
\end{center}
{\small We present a Lorentzian version of three-dimensional
noncommutative Einstein-AdS gravity by making use of the
Chern-Simons formulation of pure gravity in $2+1$ dimensions.
The deformed action contains a real, symmetric metric and a real,
antisymmetric tensor that vanishes in the commutative limit.
These fields are coupled to two abelian gauge fields.
We find that this theory of gravity is
invariant under a class of transformations
that reduce to standard diffeomorphisms once the noncommutativity parameter
is set to zero.}

\end{titlepage}

The field theory limit of open string amplitudes in the presence of a
$B_{\mu\nu}$ background gives rise to gauge theories formulated in a
noncommutative geometry \cite{Seiberg:1999vs}. The parameter
$\theta^{\mu\nu}$ which encodes
the noncommutativity properties is essentially given by $B_{\mu\nu}^{-1}$.
In a field theory perturbative framework it naturally
produces an effective cutoff, $\Delta x^\mu\equiv \theta^{\mu\nu}p_\nu$, that
modifies the UV behaviour of the theory \cite{Minwalla:1999px}.
As long as $\theta^{\mu\nu}$ is
treated as an external, given background, this amounts to an ad hoc procedure.
In order to gain true insights one would like to see
noncommutativity/nonlocality entering the game as a dynamical object. One
way to make progress in this direction might be to understand how to
formulate gravity in a noncommutative geometry.

While noncommutative gauge theories have been studied extensively at the
perturbative and nonperturbative level, not much is known about the
corresponding formulation of a gravitational theory. One of the main
obstacle to overcome is related to the fact that the Moyal product, which
implements the noncommutativity, does not maintain reality. In the
noncommutative formulation of a gauge theory this does not represent a real
problem since the deformed gauge transformations are such that they produce
new, real gauge fields.
One possible way to preserve reality of the gravitational field is to
use explicitely the Seiberg-Witten map \cite{Chamseddine:2000si}. Otherwise
it seems that one is forced to complexify the
fields \cite{Chamseddine:2000zu,Chamseddine:2000rj}.
However, complex gravity may be plagued by inconsistencies
already at the commutative level \cite{Damour:1992bt,Moffat:fc}.

In two and in three dimensions however one can take advantage of the fact
that a theory of gravity can be formulated as a gauge theory. In fact since
we know how to deform gauge transformations and since the metric does not
appear in the volume form, in order to obtain the corresponding
noncommutative version of the theory it is sufficient to introduce
the $\star$-product and extend the gauge group appropriately. 

In this letter we study noncommutative gravity in three dimensions, while
the two-dimensional theory will be treated in a forthcoming
paper \cite{cckmsz:2002}.

It is well-known that in three dimensions pure Einstein gravity can be
written as a Chern-Simons
theory. More precisely the Einstein-Hilbert action with negative cosmological
constant is equivalent to
the difference of two $SO(2,1)$ Chern-Simons actions, modulo boundary
terms \cite{Achucarro:vz,Witten:1988hc}. This was used in
\cite{Banados:2001xw}\footnote{For extensions cf.~\cite{Nair:2001kr}.}
in order to obtain a {\em Euclidean} version of
three-dimensional noncommutative gravity, based on the gauge
group $GL(2,\bC)$. We follow a similar approach and
define a three-dimensional {\em Lorentzian} version of
noncommutative gravity.  One advantage of our formulation is given by the
fact that all the fields are real and the metric is naturally identified.
We find a set of transformations which are invariances of the action and
reduce to standard diffeomorphisms in the commutative limit. Here we
present the main results; more details and insights will be given
in \cite{cckmsz:2002}.

\vspace{0.8cm}

Let us introduce the action

\begin{equation} \label{ncchernsim}
S = \beta \int {\mathrm{Tr}} \lp A \sw dA +\frac 23 A \sw A \sw A \rp
-\beta \int {\mathrm{Tr}} \lp \At \sw d\At +\frac 23 \At \sw \At \sw \At
\rp \,,
\end{equation}

where $\beta$ denotes a dimensionless coupling constant to be
specified below. The gauge fields are given by

\eqn
A &=& \omega +\frac el +\frac i2 b\,, \\
\At &=& \omega -\frac el +\frac i2 \tilde b\,,
\feqn

with

\eqn
\omega = \omega^a \tau_a \,, \qquad e = e^a \tau_a \,, \qquad b=b
\mathbb{I}
\feqn

($a = 0,1,2$) taking values in the Lie algebra u(1,1), whose generators
$\tau_A$ ($A = 0,1,2,3$)
are given in the appendix. We furthermore defined

\eqn
A \sw B \equiv A_{\mu} \star B_{\nu}\, dx^{\mu} \wedge dx^{\nu}\,,
\feqn

where the star denotes the usual Moyal product. Via
the isomorphism $so(2,1) \sim su(1,1)$, the algebra $u(1,1)$
is the natural noncommutative generalization of $so(2,1)$. Thus
we choose $U(1,1)_{\star}\times U(1,1)_{\star}$ as the gauge group
for (\ref{ncchernsim})\footnote{In order to formulate a
Lorentzian noncommutative gravity with positive cosmological constant,
one starts from the $SL(2,\bC)$ Chern-Simons action, which becomes then
$GL(2,\bC)$ in the noncommutative case, like in Euclidean
AdS gravity \cite{Banados:2001xw}. A Euclidean noncommutative
de~Sitter gravity can
instead be obtained from a $U(2) \times U(2)$ CS action.}.
An element $g\in U(1,1)_{\star}$ must satisfy the relation

\eqn
g_{\star}^{-1} = \eta g^{\dag}\eta \,,
\feqn

where $\eta = {\mathrm{diag}}(-1,1)$, and
$g \star g_{\star}^{-1} = 1 = g_{\star}^{-1}\star g$.\\
All the fields are chosen to be real and this reality is preserved under the
gauge transformations

\eqn
A \to g_{\star}^{-1} \star A \star g
        + g_{\star}^{-1} \star dg\,, \nonumber \\
\At \to \gt_{\star}^{-1} \star \At \star \gt
        + \gt_{\star}^{-1} \star d\gt\,, \nonumber
\feqn

that leave invariant (modulo boundary terms) the action (\ref{ncchernsim}).

With the definitions

\eqn
\mathcal{R} = d\omega + \omega \sw \omega \,, \\
\mathcal{T} = de + \omega \sw e + e \sw \omega \,,
\feqn

the action in (\ref{ncchernsim}) becomes

\eqn
S &=& \frac{2\beta}{l} \int {\mathrm{Tr}} \lp 2e\sw \mathcal{R} +\frac{2}{3l^2}
e \sw e \sw e \rp \nonumber \\
&& - \frac{\beta}{4} \int {\mathrm{Tr}} \lp b\sw db+\frac i3 b\sw b\sw b \rp
+ \frac{\beta}{4} \int {\mathrm{Tr}} \lp \bt\sw d\bt +\frac i3 \bt\sw \bt \sw
\bt \rp \nonumber \\
&& + \frac{i\beta}{l} \int {\mathrm{Tr}} \left\{ \lp \omega \sw e+e \sw \omega
\rp
\sw \lp b+\bt \rp \right\} \cr
&& + i\beta \int {\mathrm{Tr}} \left\{\lp \omega \sw
\omega +\frac{1}{l^2}e \sw e \rp \sw \lp b-\bt \rp \right\}\,. \nonumber
\feqn

Using ${\mathrm{Tr}}(d\omega)=0={\mathrm{Tr}}(de)$, this simplifies to

\eqn \label{nceinstein}
S &=& \frac{2\beta}{l} \int {\mathrm{Tr}} \lp 2e\sw \mathcal{R} +\frac{2}{3l^2}
e\sw e \sw e \rp \nonumber \\
&& - \frac{\beta}{4} \int {\mathrm{Tr}} \lp b\sw db+\frac i3 b\sw b\sw b \rp
+ \frac{\beta}{4} \int {\mathrm{Tr}} \lp \bt\sw d\bt +\frac i3 \bt\sw \bt \sw
\bt \rp \nonumber \\
&& + \frac{i\beta}{l} \int {\mathrm{Tr}} \left\{ \mathcal{T} \sw \lp b+\bt \rp
\right\} +i\beta \int {\mathrm{Tr}} \left\{ \lp\mathcal{R}+\frac{1}{l^2}
e\sw e \rp \sw \lp b-\bt \rp \right\}\,.
\feqn

This form of the action suggests to interpret $e^a$ as the dreibein fields
and $\omega$, $\mathcal{T}$ and $\mathcal{R}$ as connection,
torsion and curvature respectively. In fact one can easily make contact
with the usual definitions. We have

\eqn
\mathcal{R} &=& d\omega +\omega \sw \omega =d\omega^a \tau_a +\omega^a
\sw \omega^b \tau_a \tau_b \nonumber \\
&=& R^a \tau_a -\frac i2 \eta_{ab} \omega^a \sw \omega^b \tau_3\,,
\feqn

where

\eqn
R^a = d\omega^a -\frac 12 \epsilon^a_{\,\,\,bc} \omega^b \sw \omega^c\,.
\feqn

If one defines

\eqn
\omega^{ab}=\epsilon^{abc}\omega_c \,, \qquad  R^{ab}=\epsilon^{abc}R_c\,,
\feqn

then

\begin{equation}
R^{ab} = d\omega^{ab} +\frac 12 \lp \omega^a_{\,\,\, c} \sw \omega^{cb}
         - \omega^b_{\,\,\, c} \sw \omega^{ca} \rp\,.
         \label{curvature}
\end{equation}

In the commutative limit the above expression reduces to the standard curvature
in terms of the spin connection.\\
Similarly for the torsion we have

\begin{equation}
\mathcal{T} = T^a \tau_a -\frac i2 \lp \omega^a \sw e_a +e^a \sw
              \omega_a \rp \tau_3\,,
\end{equation}

with

\begin{equation}
T^a = de^a +\frac 12 \lp \omega^a_{\,\,\, b} \sw e^b +e^b \sw
      \omega_b^{\,\,\, a} \rp\,.
      \label{torsion}
\end{equation}

Using the above definitions the action in (\ref{nceinstein}) can be written as

\eqn
S &=& \frac{\beta}{l} \int \epsilon_{abc} \lp e^a \sw R^{bc} -\frac{1}{3l^2}
      e^a \sw e^b \sw e^c \rp \nonumber \\
&& - \frac{\beta}{2} \int \lp b\sw db +\frac i3 b\sw b\sw b \rp +
     \frac{\beta}{2} \int \lp \bt\sw d\bt +\frac i3 \bt\sw \bt\sw \bt \rp
     \nonumber \\
&& + \frac{i\beta}{2l} \int \eta_{ab} \lp e^a \sw \omega^b +\omega^a \sw e^b
     \rp \sw \lp b +\bt \rp \nonumber \\
&& + \frac{i\beta}{2} \int \eta_{ab} \lp \omega^a \sw \omega^b +\frac{1}{l^2}
     e^a \sw e^b \rp \sw \lp b -\bt \rp
     \label{ncactionfinal}
\feqn

Now the coupling constant
$\beta$ can be identified in terms of the three-dimensional Newton constant as

\begin{equation}
\beta = \frac{l}{16 \pi G_N}\,.
\end{equation}

The action (\ref{ncactionfinal}) represents a noncommutative version of
Einstein gravity with negative cosmological constant $\Lambda = -1/l^2$,
coupled to
a pair of $U(1)$ Chern-Simons fields\footnote{As before $h\in U(1)$ means
that $h^{\dag}=h_{\star}^{-1}$.}. In particular $b+\bt$ couples to the
torsion and $b-\bt$ couples to the curvature. Note that this coupling
is a pure noncommutative effect.
 
From (\ref{ncactionfinal}) we can compute the equations of motion

\begin{eqnarray}
\frac{\delta S}{\delta e^a}=0 &\Rightarrow& R^a+\frac{1}{2l^2}
\epsilon^a{}_{bc} e^b \sw e^c -\frac{i}{4}\big[ \omega^a \sw
(b+\tilde{b}) + (b+\tilde{b})\sw\omega^a\big] \cr &&
- \frac{i}{4l}\big[ e^a \sw (b-\tilde{b}) + (b-\tilde{b})\sw
e^a\big] = 0 \,, \cr &&\cr \frac{\delta S}{\delta \omega^a} = 0
&\Rightarrow& T^a-\frac{i}{4}\big[ e^a \sw (b+\tilde{b}) +
(b+\tilde{b})\sw e^a\big] \cr && - \frac{il}{4}\big[ \omega^a \sw
(b-\tilde{b}) + (b-\tilde{b})\sw\omega^a\big]=0 \,, \cr &&\cr
\frac{\delta S}{\delta b}=0 &\Rightarrow&
F(b)-\frac{i}{2l}\eta_{ab}\big( \omega^a \sw e^b +
e^a\sw\omega^b\big) \cr && - \frac{i}{2}\eta_{ab}\big( \omega^a \sw
\omega^b + \frac{1}{l^2} e^a\sw e^b\big) = 0 \,, \cr &&\cr
\frac{\delta S}{\delta \tilde b} = 0 &\Rightarrow&
F(\tilde{b})+\frac{i}{2l}\eta_{ab}\big( \omega^a \sw e^b +
e^a\sw\omega^b\big) \cr && - \frac i2 \eta_{ab}\big( \omega^a \sw
\omega^b + \frac{1}{l^2} e^a\sw e^b\big) = 0\,,
\end{eqnarray}

where $F(b) = db+\frac{i}{2}b\sw b,\
F(\tilde{b}) = d\tilde{b}+\frac{i}{2}\tilde{b}\sw \tilde{b}$. Again one
can easily check that in the commutative limit one is left with standard
Einstein gravity plus two decoupled $U(1)$ Chern-Simons actions.
 
A natural definition of a metric follows,

\begin{equation} \label{metric}
G_{\mu\nu} = e^a_{\mu} \star e^b_{\nu}\, \eta_{ab} =
             g_{\mu\nu} + ib_{\mu\nu}\,,
\end{equation}

where

\begin{equation}\label{smetric}
g_{\mu\nu} = \frac 12 \eta_{ab} \{e^a_{\mu}, e^b_{\nu}\}
\end{equation}

is real and symmetric, and reduces to the usual expression of the
metric in the commutative case, whereas

\begin{equation}\label{ametric}
b_{\mu\nu} = -\frac i2 \eta_{ab} [e^a_{\mu}, e^b_{\nu}]
\end{equation}

is real and antisymmetric, and vanishes for $\theta^{\mu\nu} = 0$.

We will now show that the action (\ref{ncchernsim}) is invariant
under a certain class of deformed transformations that reduce to
ordinary diffeomorphisms in the commutative case\footnote{Cf.~also
\cite{Jackiw:2001jb}.}. We recall that
for $\theta^{\mu\nu} = 0$ the variation of a one-form $A$
under an infinitesimal
diffeomorphism along an arbitrary vector field $v$ is given by the
Lie derivative

\begin{equation}
{\cal L}_v A = (d i_v + i_v d) A\,,
\end{equation}

where $i_v$ is the inner product on differential forms,
e.~g.~$i_v A = v^{\rho} A_{\rho}$ for one-forms. It is easy to show that

\begin{equation}
{\cal L}_v A = \delta_{i_v A} A + i_v DA\,, \label{diffcomm}
\end{equation}

where $DA = dA + A \wedge A$ and $\delta_{i_v A} A$ denotes the
variation of $A$ under an infinitesimal gauge transformation with
parameter $i_v A$. A natural generalization of (\ref{diffcomm})
to the noncommutative case is

\begin{equation}
\Delta_v^{\star} := \delta_{i_v^{\star} A} A + i_v^{\star} DA\,,
                    \label{ncdiff}
\end{equation}

where, for $p$-forms $\omega^p$,

\begin{equation}
i^\star_v\omega^p := \frac{1}{2(p-1)!}[v^\rho \star 
\omega^p_{\rho \mu_1 \ldots \mu_{p-1}}+\omega^p_{\rho \mu_1 \ldots \mu_{p-1}}
\star v^\rho]\, dx^{\mu_1}\wedge \ldots \wedge dx^{\mu_{p-1}}\,,
\end{equation}

and $DA = dA + A \sw A =: F$. A more detailed
discussion of the deformed transformations (\ref{ncdiff})
will be presented in \cite{cckmsz:2002}. It is clear by construction
that the transformations (\ref{ncdiff}) reduce to ordinary
diffeomorphisms in the commutative limit.

From gauge invariance of the action (\ref{ncchernsim}) and from
(\ref{ncdiff}) it follows that it is
sufficient to prove that the action is invariant (modulo boundary terms)
under the transformations

\begin{eqnarray}
\delta^\prime_v A &=& i_v^{\star} DA = i_v^{\star} F\,, \nonumber \\
\delta^\prime_v \tilde A &=& i_v^{\star} D\tilde A = i_v^{\star} \tilde F\,,
                             \nonumber
\end{eqnarray}

then also invariance under (\ref{ncdiff}) holds.
It is straightforward to show that

\eqn
\delta'_v S &=& 2 \beta \int {\mathrm{Tr}}\lp F \sw i^\star_v F \rp -
                2 \beta \int {\mathrm{Tr}}\lp \tilde F \sw i^\star_v
                \tilde F \rp \nonumber \\
            &=& -2 \beta \int {\mathrm{Tr}}\lp i^\star_v F \sw F \rp +
                2 \beta \int {\mathrm{Tr}}\lp i^\star_v \tilde F \sw
                \tilde F \rp = 0\,.
\feqn

We are now interested in how $U(1,1) \times U(1,1)$
gauge transformations act on the fields $\omega$, $e$, $b$, $\tilde b$. Let

\eqn \label{lambda}
\lambda &=& \lambda^A \tau_A =\lambda^a \tau_a +\chi \tau_3 =
(\xi^a +\alpha^a) \tau_a +\chi \tau_3\,, \\  \label{tlambda}
\tl &=& \tl^A \tau_A =\tl^a \tau_a +\tilde{\chi} \tau_3 =
(\xi^a - \alpha^a) \tau_a + \tc \tau_3
\feqn

be the generators of the infinitesimal gauge transformations, written
in terms of the parameters $\xi^a$, $\alpha^a$, and $(\chi, \tilde{\chi})$,
which refer to rotations, translations and $U(1) \times U(1)$ gauge
transformations respectively. Then

\eqn
\delta_{\lambda} A^a &=& d\lambda^a -\frac 12 \epsilon^a_{\,\,\, bc}
\lp A^b \star \lambda^c -\lambda^b \star A^c \rp \nonumber\\
&& + \frac i2 \lp A^a \star \chi -\chi \star A^a \rp +\frac i2
\lp b\star \lambda^a -\lambda^a \star b \rp\,,  \\
\delta_{\lambda} b &=& d\chi -\frac i2 \eta_{ab} \lp A^a \star\lambda^b
-\lambda^a \star A^b \rp +\frac i2 \lp b\star \chi -\chi \star b \rp\,,
\feqn

with analogous relations for $\tilde{A}^a$, $\tilde b$.
Furthermore we have

\eqn
\td_{\tl} A =\delta_{\lambda} \At = 0\,.
\feqn

In what follows it will be convenient to introduce the definitions
$\hd_{(\lambda,\tl)}:=\delta_{\lambda}+\td_{\tl}$, with
$\lambda$ and $\tl$ as in (\ref{lambda}-\ref{tlambda}), and the combined
fields $b := b_1 + b_2$ and $\bt := b_1 - b_2$.

Let us now consider separately rotations, translations and 
$U(1) \times U(1)$ gauge transformations.\\

{\bf Rotations}:

Rotations are given by setting $\alpha^a = \chi = \tilde{\chi} = 0$,
i.~e.~, by taking transformations with parameters

\begin{equation}
\lambda = \xi = \xi^a \tau_a \,, \qquad \tl =\xi =\xi^a \tau_a\,.
\end{equation}

Defining $\xi_{ab} = \epsilon_{abc} \xi^c$, we obtain for the variation
of the spin connection $\omega^{ab} = \epsilon^{abc} \omega_c$

\begin{equation}
\hd_{(\xi,\xi)}\omega^{ab} = d\xi^{ab} + \omega^{\left[a|\right.}_{\,\,\,\, c}
                             \star\xi^{\left.c|b\right]}
                             - \xi^{\left[a|\right.}_{\,\,\,\, c} \star
                             \omega^{\left.c|b\right]} +
                             \frac i2 \lp b_1 \star \xi^{ab} -\xi^{ab}
                             \star b_1 \rp\,,
\end{equation}

and for the dreibein fields

\begin{equation}
\hd_{(\xi,\xi)} e^a = -\frac 12 \lp \xi^a_{\,\,\, b} \star e^b +e^b \star
                      \xi^a_{\,\,\, b} \rp + \frac{il}{2} \lp b_2 \star \xi^a -
                      \xi^a \star b_2 \rp\,.
\end{equation}

Finally the fields $b_1$, $b_2$ transform as

\eqn
\hd_{(\xi,\xi)} b_1 &=& \frac i4 \lp \xi^a_{\,\,\, b} \star
\omega^b_{\,\,\, a} -\omega^a_{\,\,\, b} \star\xi^b_{\,\,\, a} \rp \,,\\
\hd_{(\xi,\xi)} b_2 &=& -\frac{i}{4l} \epsilon_{abc} \lp \xi^{a b} \star
e^c -e^c \star\xi^{ab} \rp\,.
\feqn

{\bf Translations}:

Translations are generated by the fields

\begin{equation}
\lambda^a =\alpha^a \,, \qquad \tl^a =-\alpha^a\,,
\end{equation}

which applied to the spin connection yield

\begin{equation}
\hd_{(\alpha,-\alpha)} \omega^{ab} =
\frac 1l\lp e^{\left[a\right.} \star \alpha^{\left.b\right]} -
\alpha^{\left[a\right.} \star e^{\left.b\right]} \rp +\frac i2 \epsilon^{abc}
\lp b_2 \star \alpha_c -\alpha_c \star b_2 \rp\,.
\end{equation}

The dreibein transforms according to

\begin{equation}
\hd_{(\alpha,-\alpha)} e^a =
ld\alpha^a + \frac l2\lp \omega^a_{\,\,\, b} \star \alpha^b - \alpha^b \star
\omega_b^{\,\,\, a} \rp +\frac{il}{2} \lp b_1 \star \alpha^a -\alpha^a \star
b_1 \rp \,,
\end{equation}

whereas the variations of the fields $b_1$, $b_2$ are given by

\eqn
\hd_{(\alpha,-\alpha)} b_1 &=& \frac{i}{2l} \lp \alpha_a \star e^a -e^a
\star \alpha_a \rp\,, \\
\hd_{(\alpha,-\alpha)} b_2 &=& \frac i4 \epsilon^{abc} \lp \omega_{ab}
\star \alpha_c -\alpha_c \star \omega_{ab} \rp \,.
\feqn

$U(1) \times U(1)$ {\bf transformations}:

$U(1) \times U(1)$ transformations are given by the parameters

\begin{equation}
\lambda = \chi \tau_3 \,, \quad \tl = \tc \tau_3 \,.
\end{equation}

If we introduce the combinations

\begin{equation}
\chi = \chi_1 +\chi_2 \,, \quad \tc= \chi_1 -\chi_2\,,
\end{equation}

the spin connection and the dreibein transform as

\eqn
\hd_{(\chi,\tc)} \omega^{ab} &=& \frac i2 \lp \omega^{ab} \star \chi_1
-\chi_1 \star \omega^{ab} \rp +\frac{i}{2l} \epsilon^{abc} \lp e_c \star
\chi_2 -\chi_2 \star e_c \rp \,, \nonumber \\
\hd_{(\chi,\tc)} e^a &=& \frac i2 \lp e^a \star \chi_1 -\chi_1 \star e^a
\rp -\frac{il}{4} \epsilon^{abc} \lp \omega_{bc} \star \chi_2 -\chi_2 \star
\omega_{bc} \rp\,,
\feqn

whereas the variations of the $U(1)$ Chern-Simons fields read

\eqn
\hd_{(\chi,\tc)} b&=&d\chi+\frac i2 \lp b\star \chi-\chi \star b \rp \,,\\
\hd_{(\chi,\tc)} \bt&=&d\tc+\frac i2 \lp \bt \star \tc-\tc \star \bt \rp\,. 
\feqn

\vspace{0.8cm}

The main goal reached in our work is the formulation of a noncommutative
version of Lorentzian Einstein-AdS gravity in three dimensions
in terms of real fields.
This has been achieved in two steps: first we have used the Chern-Simons
gauge theory formulation of gravity and the fact that gauge transformations
maintain the reality properties of the fields even in the noncommutative case.
Secondly we have defined the gravity fields in such a way that the
identification of a real, symmetric metric and a real antisymmetric tensor
naturally arise. We have found that this deformed Einstein action possesses
a class of invariances that reduce to standard diffeomorphisms in the
commutative limit. We have computed how all the relevant fields transform
under rotations, translations and $U(1) \times U(1)$ gauge transformations.
An open question is still how to quantize
the theory and understand which is the dynamical role played by the
antisymmetric $b_{\mu\nu}$ field. In order to complete this program one
would need to obtain a noncommutative version of general coordinate
transformations. Our deformed transformations (which are active)
may be related to a subset of them, i.~e.~, to
the ones that maintain the Poisson structure introduced by $\theta^{\mu\nu}$.
Another interesting issue to address is how to couple gravity to matter fields.

\bigskip 
{\bf Acknowledgements}
\small

This work was partially supported by INFN, MURST and
by the European Commission RTN program
HPRN-CT-2000-00131, in which S.~C.~, D.~K.~, L.~M.~and D.~Z.~are
associated to the University of Torino.

\newpage

\normalsize

\begin{appendix}

\section{Conventions}

An element $M$ of the Lie algebra u(1,1) satisfies

\begin{equation}
M^i_{\,\,\,j} = - \eta_{jk}\bar{M}^k_{\,\,\,l}\eta^{li}\,, \label{u11}
\end{equation}

where a bar denotes complex conjugation, and
$(\eta_{ij}) = {\mathrm{diag}}(-1,1)$. We choose as u(1,1) generators

\begin{eqnarray}
\tau_0 &=& \frac 12\left(\begin{array}{cc} i & 0 \\ 0 & -i \end{array}
           \right)\,, \qquad
\tau_1 = \frac 12\left(\begin{array}{cc} 0 & 1 \\ 1 & 0 \end{array}\right)\,,
         \nonumber \\ && \nonumber \\
\tau_2 &=& \frac 12\left(\begin{array}{cc} 0 & -i \\ i & 0 \end{array}
           \right)\,, \qquad
\tau_3 = \frac 12\left(\begin{array}{cc} i & 0 \\ 0 & i \end{array}\right)\,.
         \label{generators}
\end{eqnarray}

They are normalized according to

\begin{equation}
{\mathrm{Tr}}(\tau_A\tau_B) = \frac 12 \eta_{AB}\,,
\end{equation}

where $(\eta_{AB}) = {\mathrm{diag}}(-1,1,1,-1)$ is the inner product
on the Lie algebra. The generators (\ref{generators}) satisfy the relation
(\ref{u11}). Further, if $a,b,c$ assume the values $0,1$ and $2$, then
the following relations hold:

\begin{eqnarray}
\left[ \tau_a ,\tau_b \right] &=& -\epsilon_{abc}\tau^c\,, \\
\left[ \tau_a ,\tau_3 \right] &=& 0\,, \\
\tau_a \tau_b &=& -\frac 12 \epsilon_{abc}\tau^c -\frac i2
\eta_{ab} \tau_3\,, \\
{\mathrm{Tr}}(\tau_a\tau_b\tau_c) &=& -\frac 14 \epsilon_{abc}\,, \\
{\mathrm{Tr}}(\tau_a\tau_b\tau_3) &=& \frac i4 \eta_{ab}\,,
\end{eqnarray}

where $(\eta_{ab}) = {\mathrm{diag}}(-1,1,1)$.\\
Furthermore we defined $\epsilon_{012} = 1$.

\end{appendix}

\end{document}